\documentclass[useAMS,usenatbib]{mn2e}

\usepackage{mathptmx,subfigure,graphicx,lpic, color}

\newcommand\aj{{AJ}}%
\newcommand\apj{{ApJ}}%
\newcommand\apjl{{ApJ}}%
\newcommand\apjs{{ApJS}}%
\newcommand\ao{{Appl.~Opt.}}%
\newcommand\aap{{A\&A}}%
\newcommand\aaps{{A\&AS}}%
\newcommand\mnras{{MNRAS}}%
\newcommand\procspie{{Proc.~SPIE}}%

\title[Supernova rate from SDSS-I]{The Rate of Type Ia Supernovae at
  $z \approx 0.2$ from SDSS-I
Overlapping Fields}

\author[Horesh et al.]
{A.~Horesh$,^{1}$
D.~Poznanski,$^{2}$
E.~O.~Ofek$^{3}$ and
D.~Maoz$^{1,4}$\\
\\
$^{1}$School of Physics and Astronomy and Wise Observatory, Tel Aviv University, Tel
Aviv 69978, Israel\\
$^{2}$Department of Astronomy, University of California, Berkeley, CA 94720-3411, USA\\
$^{3}$Division of Physics, Mathematics and Astronomy, California
Institute of Technology, Pasadena, CA 91125, USA\\
$^{4}$Osservatorio Astrofisico di Arcetri, Largo Enrico Fermi 5,
Firenze 50125, Italy\\}

\begin{document}

\maketitle
\label{firstpage}
\begin{abstract}

  In the course of the Sloan Digital Sky Survey (SDSS-I), a large
  fraction of the surveyed area was observed more than once due to
  field tiling overlap, usually at different epochs. We utilize some
  of these data to perform a supernova (SN) survey at a mean
  redshift of $z=0.2$. Our archival search, in $\sim 5 \%$ of the
  SDSS-I overlap area, produces $29$ SN candidates clearly associated
  with host galaxies.  Using the Bayesian photometric classification
  algorithm of Poznanski et al., and correcting for classification
  bias, we find $17$ of the $29$ candidates are likely Type Ia SNe.
  Accounting for the detection efficiency of the survey and for host
  extinction, this implies a Type Ia SN rate of $r_{\rm
    Ia}=\left(14.0^{+2.5+1.4}_{-2.5-1.1}\pm 2.5\right)\times
  10^{-14}~h_{70}^{2}~{\rm yr}^{-1}~L^{-1}_{\odot,g}$, where the
  errors are Poisson error, systematic detection efficiency error, and
  systematic classification error, respectively. The volumetric rate
  is $R_{\rm Ia}=\left(1.89^{+0.42+0.18}_{-0.34-0.15}\pm
    0.42\right)\times 10^{-5}~{\rm yr}^{-1}~h_{70}^{3}~{\rm
    Mpc}^{-3}$. Our measurement is consistent with
  other rate measurements at low redshift. An order of magnitude
  increase in the number of SNe is possible by analyzing the full
  SDSS-I database.

\end{abstract}


\section{Introduction}

Supernovae (SNe) play a central role in galaxy evolution and cosmic
metal production. Measuring the rates at which SNe explode is thus an
important step for understanding the chemical evolution of the
universe.  In recent years, efforts have intensified to measure the
low-redshift Type Ia SN rate both in field environments (Cappellaro et
al. 1999; Hardin et al.  2000; Madgwick et al. 2003; Blanc et al. 2004,
Botticella et al. 2008, Dilday et al. 2008) and in galaxy clusters
(Gal-Yam et al. 2002; Maoz \& Gal-Yam 2004; Sharon et al. 2007;
Mannucci et al. 2008; Graham et al. 2008; Sand et al. 2008).  
However, due to small SN numbers, there are
still significant uncertainties in low-redshift SN rates.

In this paper, we demonstrate that a large archival repository of SNe,
one that is potentially useful for a low-redshift rate measurement
using a large number of SNe, exists in the data from the first phase
of the Sloan Digital Sky Survey (SDSS; York et al. 2000). We use a
small fraction of these data to detect and compile a sample of SNe Ia
and to derive the SN Ia rate at low-redshift. The techniques we use in
this paper are also of relevance for future projects such as the
 Panoramic
Survey Telescope and Rapid Response System (Pan-STARRS; Kaiser 2004)
and the Large Synoptic Survey Telescope (LSST; Tyson 2002).
These projects will survey huge areas in a relatively short time, and
will produce large samples of SNe for which spectral
classification will not be possible, due to their large numbers.

In \S2 we describe the SDSS data we use.  The pipeline used to process
these data and detect SNe is presented in \S 3.  Detection efficiency
and photometric calibration are discussed in sections $4$ and $5$. \S
$6$ presents our preliminary results, including a first SN sample, its
classification, and a calculation of the SN Ia rate. We compare our
results to previous measurements in \S7, and summarize in
\S 8.

\section{SDSS Imaging Data}

The SDSS imaged about one quarter of the sky in five bands ($u$, $g$,
$r$, $i$, $z$, centered at 3551\AA, 4686 \AA, 6165 \AA, 7481 \AA, 8931
\AA; Fukugita et al. 1996). Images were photometrically (Tucker et al.
2006) and astrometrically (Pier 2003) calibrated by the SDSS pipeline
(Lupton et al. 2001). The data products of the SDSS (images and object
catalogs) were made available\footnote{http://www.sdss.org} in a
series of Data Releases (see Adelman-McCarthy et al. 2008 for a
description of the latest data release, DR6).

With the objective of covering the survey area once, imaging was
performed by scanning the sky in great circles. Each scan was along a
$2.5$-degree-wide strip, where each strip was divided into numerous
``fields''. However, dividing the celestial sphere on to rectangular
planes causes the rectangles to overlap, especially close to the poles
of the survey scan coordinate system (see e.g. Fig. 1). In addition,
adjacent strips have an intentional overlap for the purpose of
photometric and astrometric quality checks. The fact that different
strips were imaged at different times raises the possibility of using
the overlap regions to detect transient events.
\setcounter{figure}{0}
\begin{figure}
\includegraphics[width=0.48\textwidth]{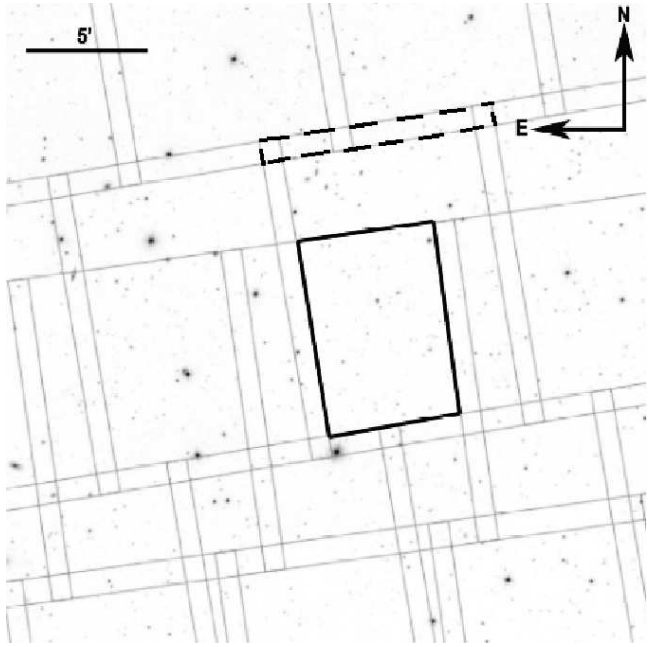}
\caption{Examples of overlap between SDSS-I fields centered around
  RA=14:56:49, Dec=+10:57:48. The dashed polygon is an overlap between
  two adjacent strips which is used for SDSS quality checks and the
  solid polygon is an overlap resulting from the mapping of the
  celestial sphere on to rectangles.}
\end{figure}

Each final SDSS field is an image of $2048\times1361$ pixels. The
image pixel scale is $0\farcs 396$ with a median point spread function
(PSF) of $1\farcs 4$ in the $r$ band. An exposure time of $53.9$ sec
was used to image all fields, resulting in point source AB magnitude
$95\%$ repeatability limits of $22.0,~22.2,~22.2,~21.3$ and $20.5$, in
the $u,~g,~r,~i,~z$ bands, respectively.

\section{Supernova Survey Data Pipeline}

To deal with the vast amounts of data in the SDSS database, we
developed a largely automatic pipeline for downloading individual
subsets of overlapping field images, and processing them one at a
time. Our pipeline consists of three independent modules for download,
registration, and detection, executed in that order.

As a first step, we compiled a list of overlapping SDSS fields. To do
so, we downloaded the coordinates of all the fields in the SDSS DR4
database. By applying a polygon intersection algorithm, which assumes
planar geometry, to the list of coordinates, we constructed a list of
the overlapping regions of each SDSS field. Each image set, consisting
of a first-epoch image (the ``reference'' epoch) and its overlapping
second epoch images, was individually downloaded for further
processing by our pipeline. In the present paper, we search for SNe in
the region $220^{\circ}<{\rm RA}<240^{\circ}$, and $-1^{\circ}<{\rm
  Dec}<64^{\circ}$. This region is not far from the pole of the SDSS
coordinate system (${\rm RA}=275^{\circ}$, ${\rm Dec}=0^{\circ}$),
resulting in a large overlap area of $92~{\rm deg}^{2}$, obtained from
$460~{\rm deg}^{2}$ of SDSS images.

Our survey search method is based on image subtraction. We note that
an alternative method is to search for SNe in the SDSS catalog using
different criteria, e.g., SN colors (see Poznanski et al. 2002).
However, a SDSS catalog SN search has some disadvantages. For example,
a blind color search will be affected by color contamination
originating from SN host galaxies. In addition, lacking a direct
access to the SDSS pipeline makes it difficult to estimate the survey
detection efficiency function.

We chose to limit our SN search to the $g$ and $r$ bands since they
are the deepest bands in the SDSS. Furthermore, the scanning order of
each field in the SDSS is $r,~i,~u,~z,~g$. Therefore, the $r$- and
$g$-band exposures of the same field in a given scan have the largest
time separation, i.e., there is a $\sim 5$ minute difference between
exposures of the same field in these bands. This time difference is
critical for identifying and excluding solar-system objects from among
the SN candidates.

Two computers were used for running our pipeline. One computer was
used for continuous downloading of images from the SDSS database. In
parallel, the registration and detection modules (see below) were run
on a computer with a Pentium IV 3.4 GHz processor and 2 GB of memory.
The download rate and the processing rate both dictated a net data
flow rate of about 1 deg$^2$ per day. In practice, software, hardware,
and communications problems resulted in a lower rate, and guided our
decision to stop the current search after about 90 deg$^2$.

\subsection{Image registration}

The registration module aligns the overlapping images in each set to
their reference image and produces a difference image in which SN
candidates are searched for by the detection module. Both the $g$ and
$r$ image sets, once downloaded, are registered separately by the
registration module. An overlapping image is first registered to its
reference image, based on their world coordinate system (WCS)
coordinates, using the $wregister$ IRAF (Tody 1986) task. The
overlapping areas are then cut out from both the reference and
registered images.  Using each field's photometric parameters,
extracted from the SDSS
database\footnote{http://casjobs.sdss.org/casjobs}, we next linearly
match the background level and the zeropoint of the registered image
to those of the reference image. At this point, we try to achieve a
better image registration, which is crucial for image subtraction, by
matching the positions of objects which appear in both images. We
first detect objects in the overlapping image segments by applying the
Source Extractor (SExtractor) program (Bertin \& Arnout 1996) to both
segments. Next, by cross correlating the object positions, a more
precise registration between the two segments is obtained using the
$geomap$ and $geotran$ IRAF tasks, allowing for offsets in the $x$ and
$y$ axes and a rotation angle between the two images. In order to
avoid poor statistics in the matching process, this latter alignment
is performed only if there are at least seven matching objects.
Otherwise this stage is skipped.

Next, the image with the smaller PSF full width at half maximum (FWHM)
is degraded by convolving it with a ${\rm 2D}$ Gaussian kernel,
$G(x,y)\propto {\rm exp}[-(x^2+y^2)/2\sigma^2]$, in order to match the
PSF of the second image. The kernel is found from the parameters of
the two image PSFs, listed as ``psfWidth'' in the SDSS catalogue
``Field'' Table. This simplistic PSF matching approach is dictated by
the small number of objects (generally not point sources) in the
overlapping regions, which prevents the application of more
sophisticated PSF-matching algorithms (e.g. Alard \& Lupton 1998;
Alard 2000).

Following subtraction of two registered images, the absolute values of
the difference image is formed, so that all residuals are positive. In
order to smooth out residuals due to imperfect alignment, the
difference image is smoothed by convolving it with a 2D Gaussian,
three pixels wide $(1\sigma)$.

\subsection{SN candidate detection}

The residuals are detected in the difference image by applying
SExtractor to the image. Since the final difference image is positive
definite (see $\S3.1$), it has a one sided noise distribution. We
chose to apply a $6\sigma$ detection threshold in the detection
process. The value of the detection threshold is calculated using the
Poisson fluctuations of the background counts in both the reference
and registered images.  The residuals detected by SExtractor are
automatically examined in more detail to screen for various non-SN
detection contaminations, as described next.

We first search for variable stars within our candidate list. Using
SExtractor, we obtain a list of objects in both the registered and
reference images. If an object is detected in both images at the same
position where a residual was detected in the difference image, we
query the SDSS catalogue for objects at that position. If an object,
catalogued as a star, exists at that position, the candidate is
considered a variable star and is rejected from our candidate list.
Similarly, we reject candidates spectroscopically identified as
quasars.

We next explore the possibility that a residual is the result of poor
image registration. We search, using SExtractor, for positive
residuals in two new difference images: the reference minus registered
image, and vice versa. If in each of the two images a residual is
detected near the position of a candidate, we compare the difference
between the photon counts of the two residuals to our detection
threshold. In contrast to our original detection in the absolute value
of the difference image, we now require the photon count difference to
have at least $3\sigma$ significance. In a final test for improper
alignment, a stamp of $41\times41$ pixels around the position of each
residual is cut out of the registered and reference images. The two
image stamps are re-registered using the $xregister$ task in IRAF,
based on cross correlation. A new difference image is produced using
the new re-registered images. If no residuals are detected in the new
subtracted stamp image, the candidate is discarded.

The remaining candidates are subjected to another test, aimed at
determining whether or not they are moving objects. We first query the
SDSS catalogue to check if the target has been flagged as a moving
object. We also compare the position of each candidate in the $g$ and
$r$ bands, assuming that it was detected in both bands. If the
candidate position has changed by more than $2$ pixels, it is also
considered a moving object and is excluded from the candidate list.
All stages up to this point are performed automatically, with no human
intervention.

The remaining candidates are saved, together with their subtraction
images, for visual inspection, performed by a single person (AH). The
inspection helps reject false positives of various types, such as
artifacts and 
residuals due to poor PSF matching, poor image alignment, cosmic rays,
and saturated objects (see Fig. 2). About 99\% of the candidates found by the
automatic pipeline are discarded as false positives in the visual
inspection stage.
\setcounter{figure}{1}
\begin{figure}
\centering
\includegraphics[width=0.21\textwidth, height=0.2\textwidth]{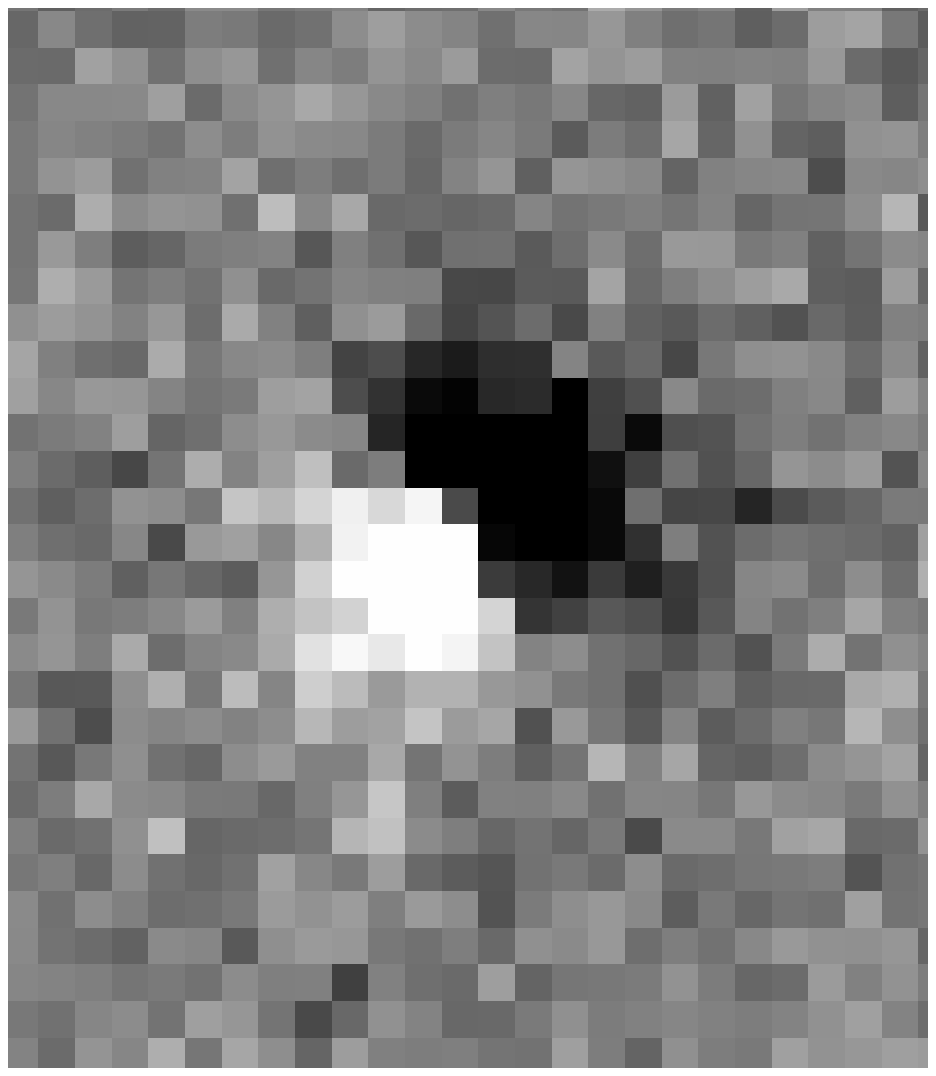}
\includegraphics[width=0.2\textwidth, height=0.2\textwidth]{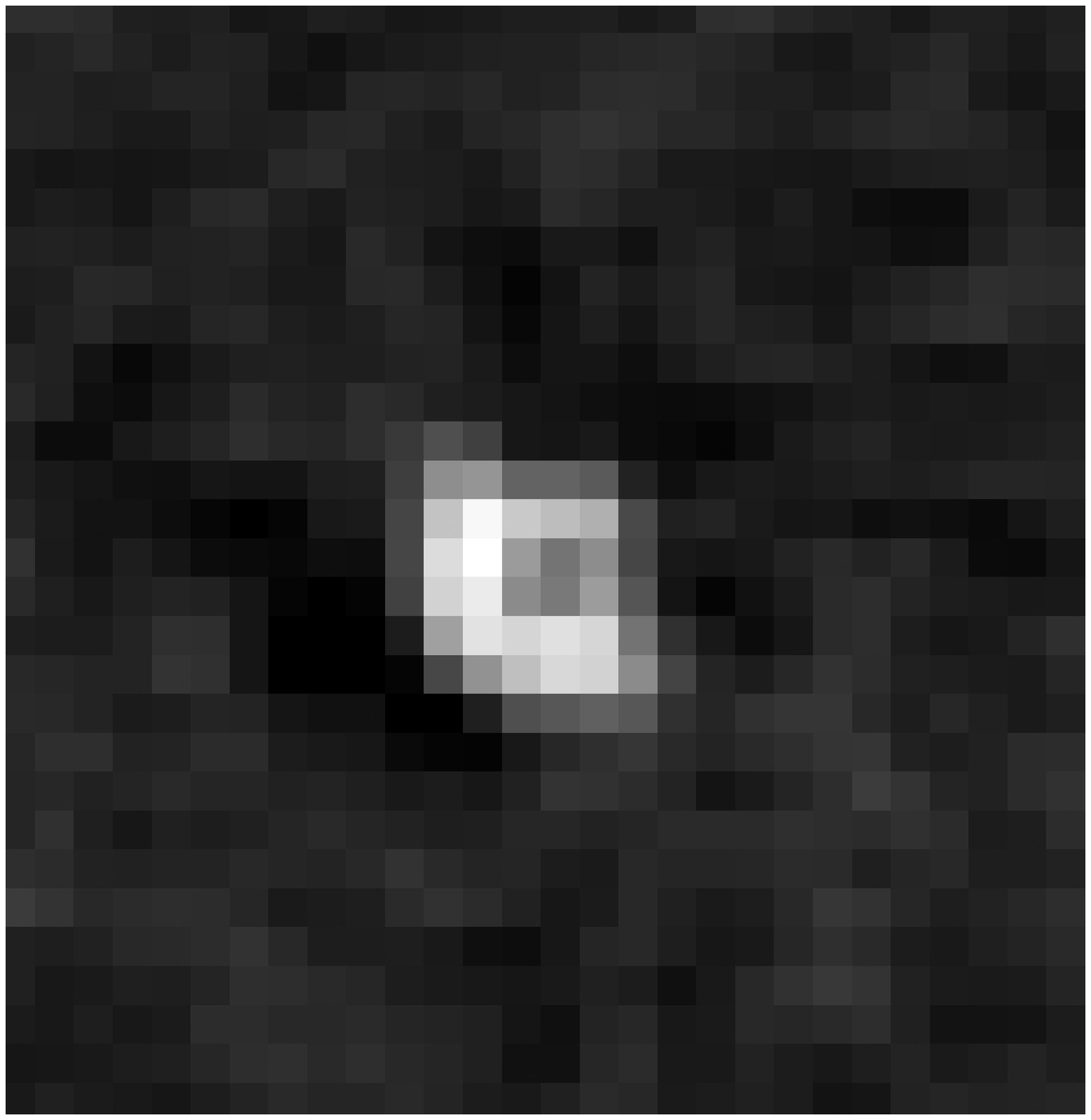}
\caption{Examples of false positives which were rejected in
  the visual inspection stage. The left panel shows a residual in the
  subtraction image due to image misalignment. The residual in the
  right panel is due to poor PSF matching.}
\end{figure}

\section{Detection efficiency functions}

Estimating the efficiency of our SN detection process is critical for
deriving a reliable SN rate. 
We have planted a sample of fake SNe in the SDSS images,
whose recovery fraction provides an estimate of the detection
efficiency as a function of SN magnitude. 

The first step in producing the fake SN sample was choosing the SN
hosts. We compiled the $g$ magnitude and the photometric redshift
(Csabai et al. 2003; Oyaizu et al. 2008) of each galaxy in the
overlapping sets of images from the SDSS catalog.  Under a simplifying
assumption that the SN rate is proportional to stellar $g$-band
luminosity, we selected a random subset of galaxies weighted by
luminosity. To each of these selected hosts we then assigned a fake
SN. The SN was assigned a random $g$-band absolute magnitude in the
range of $-19.5$ to $-7.5$. The absolute magnitudes were converted to
observed magnitudes using a distance modulus based on the SN host
photometric redshift (assuming a Hubble parameter of $H_{0}=70~{\rm km
  ~s}^{-1}{\rm Mpc}^{-1}$, a mass density in units of the critical
density $\Omega_{\rm m}=0.3$, and cosmological constant
$\Omega_{\Lambda}=0.7$). The SN $r$-band magnitude was randomly chosen
to be in the range of $-0.5$ and $+1.5$ of the $g$ band magnitude, a
range motivated by calculating synthetic $g$ and $r$ magnitudes from a
set of observed spectra of SNe Ia (Nugent 2002; Poznanski et al. 2002;
Poznanski, Maoz, \& Gal-Yam 2007).

The fake SNe were added to the real images as part of the data
processing, prior to image registration, as follows. First, we
randomly chose the image in which the SN was to be planted, i.e.,
either the reference image or the registered one. We then cut out a
region around the fake SN host of size $1.5$ times the host's $90\%$
light radius. SExtractor was applied to the host stamp image,
producing a list of $10\%,~20\% ,...,~100\%$ light radii of the host.
The radial distance of the fake SN from its host center was chosen
randomly from among these annuli, assuring that the locations of the
artificial SNe roughly follow their galaxy host light. The final
position of the fake SN with respect to the host was at a randomly
chosen position angle. The SN was then planted in the selected image
using the IRAF task $mkobjects$.

The fake SN sample underwent the same processing as the real data,
including the visual inspection stage, ensuring it reflected
faithfully the actual detection efficiency. By spreading the fake SNe
among all the overlapping fields, we also took into account the fact
that the efficiency may vary from field to field.

Our detection efficiency functions in the $g$ and $r$ bands are shown
in Figure $3$. We find that our efficiency level is $\sim 60\%$, at
best. This is probably due to the poor quality of image subtraction
when the PSF matching and/or the registration are not perfect. For
example, bright hosts often leave large residuals at their centers in
the difference images. Therefore, a real SN that is close to its host
center, may be mistaken for an artificial residual due to poor
registration, by either the pipeline or the human inspector. This
effect seems to be independent of magnitude, even for bright SNe, due
to the fact that such SNe will tend to be hosted by nearby, and hence
bright, galaxies. It is also evident that the detection efficiency in
the $r$ band starts declining at brighter magnitudes, compared to the
$g$ band, probably due to the relative faintness of galaxies in the
$g$-band.
\setcounter{figure}{2}
\begin{figure}
  \centering
  \subfigure{\includegraphics[width=0.5\textwidth]{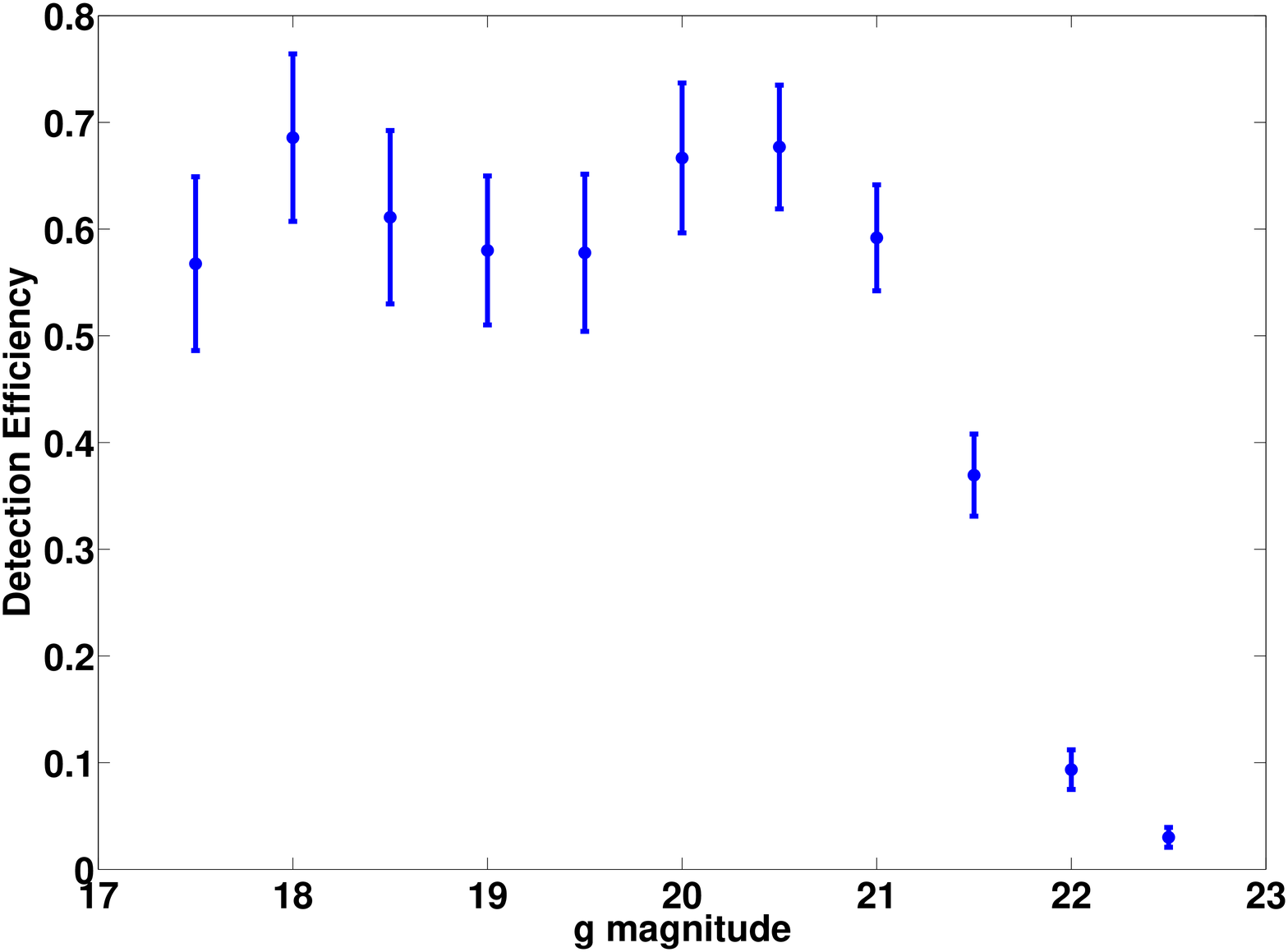}}

  \subfigure{\includegraphics[width=0.5\textwidth]{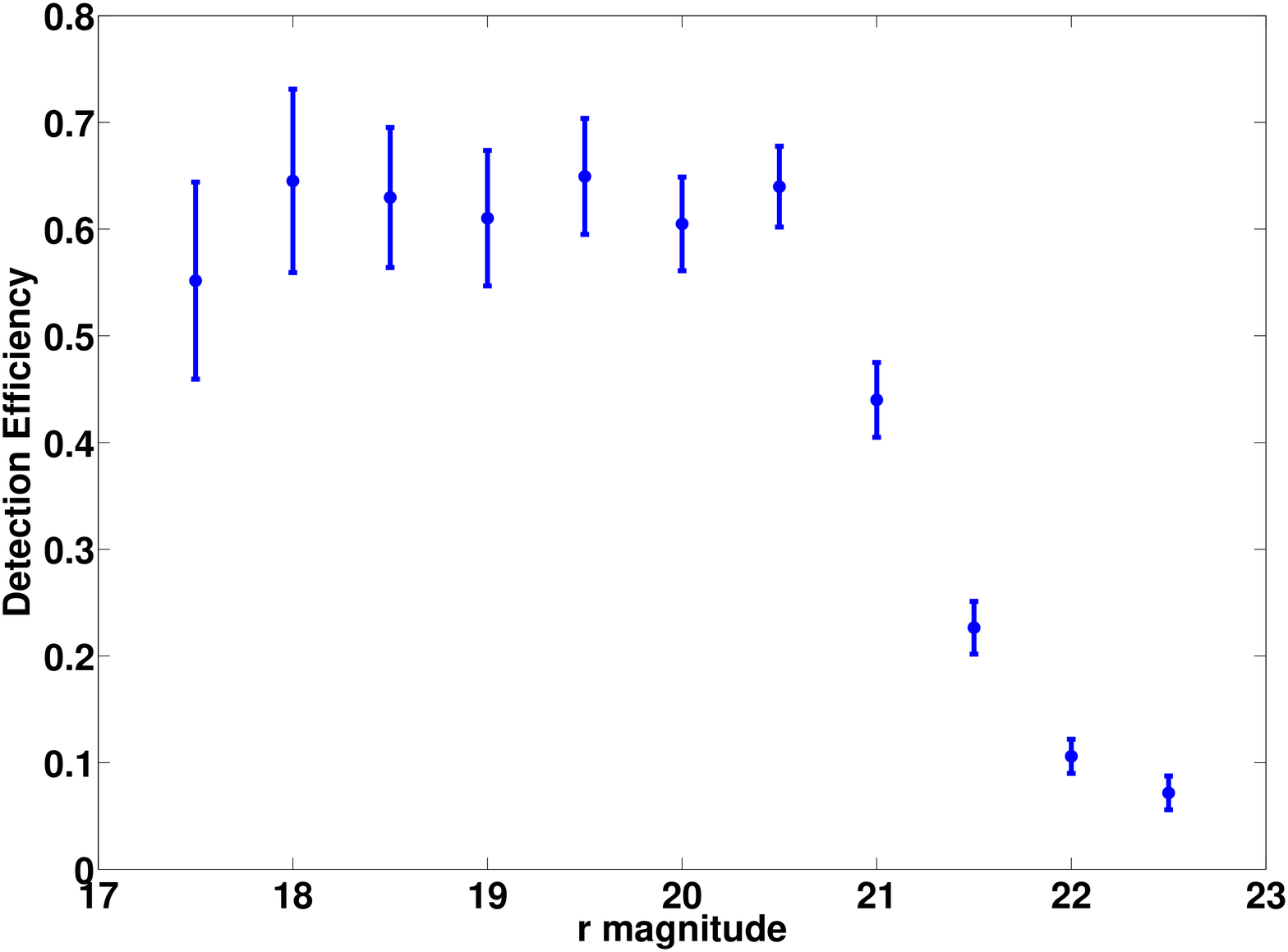}}
\caption{Detection efficiency functions in the $g$ (top) and $r$ (bottom) bands. Error bars represent
  $1\sigma$ Poisson Errors.}
\end{figure}

\section{Photometric Calibration}

The SN candidates which pass the visual inspection are further
explored. In order to obtain the magnitudes of these candidates, we
download the images, in which a candidate resides, in the remaining
($u$, $i$, $z$) SDSS bands. Since the images of a field in different
bands do not fully overlap, we first align them according to the image
in which the candidate was detected (in either the $g$ or the $r$
band). We perform this alignment for the reference field images and
for the registered field images separately. Then, the reference and
registered field images in each band are processed by the same
registration module used to originally process the $g$ and $r$ bands
(see $\S3$).  After a final difference image is obtained in all five
bands, $41 \times 41$-pixel images are cut out around the candidate.

These images are used to perform aperture photometry of the
candidates. The counts are summed in an aperture of radius $2\sigma$,
where $\sigma$ is $1/2.35$ of the FWHM of the average PSF of a field
(taken from the SDSS catalog ``Field'' table). With the zeropoint,
airmass, and extinction parameters, also listed in the SDSS catalog,
we convert the candidate counts to magnitudes. To each magnitude we
then apply an aperture correction, that accounts for the flux outside
the aperture.  From aperture photometry on bright and isolated SDSS
stars, we find mean aperture corrections of $0.21, 0.22, 0.23, 0.25,
0.25$ in the $u,~g,~r,~i,~z$ bands, respectively. With these
corrections, our final magnitudes for bright stars also match those in
the SDSS catalogue.  We corrected the candidate magnitudes for
Galactic extinction according to Schlegel et al. (1998).

To obtain realistic error estimates for our magnitude measurements, we
have performed the same analysis on a large sample of artificial SNe.
The artificial SNe were blindly planted in several images in the same
manner as done for obtaining our detection efficiency functions (see
$\S4$). We expect the photometric errors to be larger than the usual
Poisson errors due to inaccurate registration, varying backgrounds,
and residuals from the host galaxy subtraction. We planted $\sim 400$
artificial SNe in each of eight magnitude bins (see
Table 1). In each bin, and for each band, we calculate the root-mean
square (rms) of the difference between the measured magnitudes and the
original magnitudes assigned to the artificial SNe, and adopt it as
the systematic photometric error. The results are listed in Table~1.
\begin{table}
\begin{center}
\caption{Photometric Errors}
\smallskip
\begin{tabular}{ccccccccc}
\hline
\noalign{\smallskip}
Band & \multicolumn{8}{c}{$1 \sigma$ Magnitude Errors}\\
     & $19$ & $19.5$ & $20$ & $20.5$ & $21$ &
     $21.5$ & $22$ & $22.5$\\
\cline{2-9}
\noalign{\smallskip}

\noalign{\smallskip}
$u$  & $0.10$ & $0.11$ & $0.17$ & $0.16$ & $0.17$ & $0.26$ & $0.36$ & $0.45$ \\
$g$  & $0.10$ & $0.13$ & $0.13$ & $0.12$ & $0.17$ & $0.17$ & $0.24$ & $0.24$ \\
$r$  & $0.11$ & $0.11$ & $0.17$ & $0.13$ & $0.15$ & $0.32$ & $0.30$ & $0.29$ \\
$i$  & $0.11$ & $0.14$ & $0.16$ & $0.22$ & $0.24$ & $0.63$ & $0.55$ & $0.50$ \\
$z$  & $0.18$ & $0.19$ & $0.26$ & $0.60$ & $0.62$ & $0.69$ & $0.86$ & $0.61$ \\

\noalign{\smallskip}
\hline
\end{tabular}
\end{center}
\end{table}

\section{Results}

\subsection{SN sample}

Our final candidate list for the $92~{\rm deg }^{2}$ of overlap area
searched includes $47$ transient candidates which we denote as
SISN01 to 47, where SISN stands for SDSS-I SN. Among the candidates,
$25$ are clearly associated with a detected host galaxy, but offset
from the nucleus (if there is one). A further $11$ are, to within SDSS
resolution, at the centers of their hosts. A final $11$ candidates are
``hostless'', i.e., cannot be unambiguously associated with any
detected galaxy. Our criterion for hostlessness is
 being separated by both $>5''$ and more than two times the
$90\%$ light radius from any galaxy.

We first turn our attention to estimate the sample contamination by SN
``impostors''. Based on the SDSS limiting magnitudes (see section $\S
2$), we estimate that a true SN Ia, at maximum light, will be
undetectable at redshifts $z>0.35$. Although there are exceptions,
most core-collapse SNe are less luminous than SNe-Ia. Candidates with
spectroscopic or photometric host redshifts with a $1\sigma$ lower
limit above $z>0.35$ are therefore excluded. Six of the seven
candidates excluded by this criteria are at their host centers, and
are thus likely to be active galactic nuclei (AGN), rather than SNe.
The seventh candidate, SISN$47$, which is not at the center of its host,
might be a valid SN candidate falsely rejected due to an error in the
photometric redshift of its host. However, if it were
included in our final sample it would have no effect on the total number of
Type Ia SNe, being classified as a likely core-collapse SN (see Table 2).

The hostless candidates could be either real SNe which reside in
galaxies below the SDSS limiting magnitude, or they can be impostors
such as quasars, slow-moving asteroids, and variable stars. To
estimate the expected fraction of SNe hosted by galaxies fainter than
the SDSS limiting magnitude, we use the Blanton et al.  (2003) galaxy
luminosity function in the $r$ band. The fraction of the stellar
luminosity in galaxies with a luminosity $L<L_{{\rm lim}}$ is
\begin{equation}
P(L<L_{{\rm lim}})=\frac{\int\limits_{0}^{L_{\lim}}L\phi(L)dL}{\int\limits_{0}^{\infty}L\phi(L)dL},
\end{equation}
where $\phi$ is the luminosity function. At $z=0.2$, the mean redshift
of SNe probed by our search (see $\S 6.3$, below), the SDSS $r$-band
flux limit correspond to an absolute magnitude of $M_{r}=-17.8$ mag,
and $P(L<L_{{\rm lim}})=23\%$. Again assuming that SNe track the
stellar luminosity, we therefore expect $23\% \times40 \approx9$
candidates in galaxies below the SDSS limiting magnitude, consistent
with the $11$ hostless candidates we find. Conversely, this also
argues that most of the hostless candidates are likely real SNe, as
otherwise a large deficit of SNe in low-luminosity galaxies would be
implied.  Nonetheless, due to the lack of redshifts for the hostless
candidates, we are unable to determine with great confidence which of
those candidates are real SNe and what are their types. We therefore
exclude the $11$ hostless candidates from our sample for the purpose
of the SN rate calculation.  The exclusion of the hostless candidates
is accounted for in the luminosity-normalized SN rate calculation by
using the luminosity density which originates from galaxies above the
SDSS limiting magnitude (see $\S6.3$).

Kuiper belt objects (KBOs) might also play a role as SN impostors.
However, they are normally found near the ecliptic, while all our SN
candidate have ecliptic latitude $\beta >15^{\circ}$. Moreover,
according to their magnitude distribution (Bernstein et al. 2004)
their expected number in our survey, even near the ecliptic, is at
least an order of magnitude lower than the observed number of SN
candidates. We have checked for asteroids near the positions of all
candidates in the Minor Planet
Center\footnote{http://www.cfa.harvard.edu/iau/mpc.html} and Nasa Jet
Propulsion Laboratory\footnote{http://ssd.jpl.nasa.gov/sbfind.cgi}
databases. No known asteroids were found within $5'$ of any
candidates. Our final main SN sample thus consists of $29$ candidates
with hosts, among which five are nuclear and therefore may be AGNs.
Figures $4,~5,~6$ show sections of the reference, registered, and
difference images for each candidate in the final, hostless and
high-$z$ samples, respectively.  
\setcounter{figure}{3}
\begin{figure}
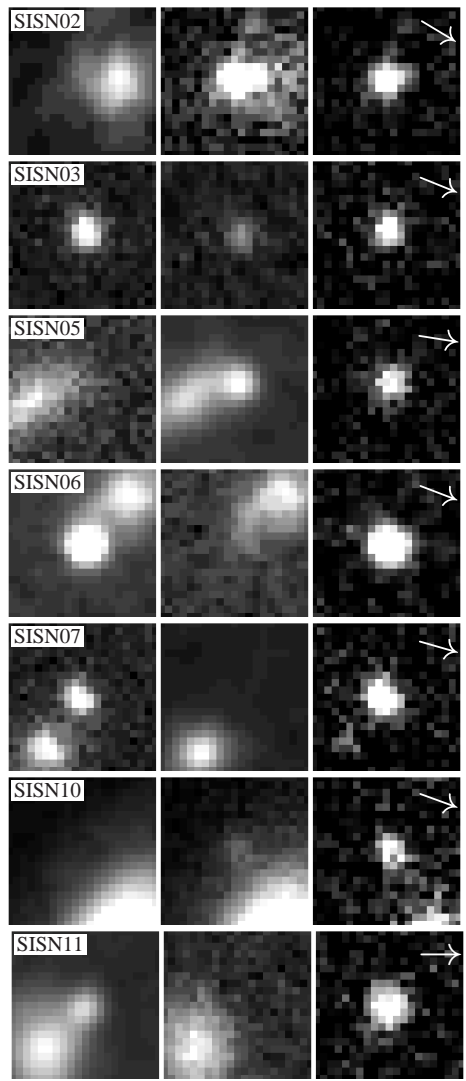

\centering
\begin{lpic}{f4a(0.35)}
\setlength{\lpbgboxsep}{0.2mm}
\lbl[W]{15.5,51;SISN02}
\lbl{164,47,-121;{\huge{\color{white}$ \uparrow$}}}
\end{lpic}
\begin{lpic}{f4b(0.35)}
\setlength{\lpbgboxsep}{0.2mm}
\lbl[W]{15.5,51;SISN03}
\lbl{164,47,-112;{\huge{\color{white}$ \uparrow$}}}
\end{lpic}
\begin{lpic}{f4c(0.35)}
\setlength{\lpbgboxsep}{0.2mm}
\lbl[W]{15.5,51;SISN05}
\lbl{164,47,-99;{\huge{\color{white}$ \uparrow$}}}
\end{lpic}
\begin{lpic}{f4d(0.35)}
\setlength{\lpbgboxsep}{0.2mm}
\lbl[W]{15.5,51;SISN06}
\lbl{164,47,-111;{\huge{\color{white}$ \uparrow$}}}
\end{lpic}
\begin{lpic}{f4e(0.35)}
\setlength{\lpbgboxsep}{0.2mm}
\lbl[W]{15.5,51;SISN07}
\lbl{164,47,-106;{\huge{\color{white}$ \uparrow$}}}
\end{lpic}
\begin{lpic}{f4f(0.35)}
\setlength{\lpbgboxsep}{0.2mm}
\lbl[W]{15.5,51;SISN10}
\lbl{164,47,-110;{\huge{\color{white}$ \uparrow$}}}
\end{lpic}
\begin{lpic}{f4g(0.35)}
\setlength{\lpbgboxsep}{0.2mm}
\lbl[W]{15.5,51;SISN11}
\lbl{164,47,-90;{\huge{\color{white}$ \uparrow$}}}
\end{lpic}


\caption{Final sample of SN candidates - For each candidate the reference (left),
  registered (center), and difference (right) images are shown. Images
  are $16''$ on a side. Arrows indicate north, with east to the left
  when facing north.}
\end{figure}

\setcounter{figure}{3}
\begin{figure*}
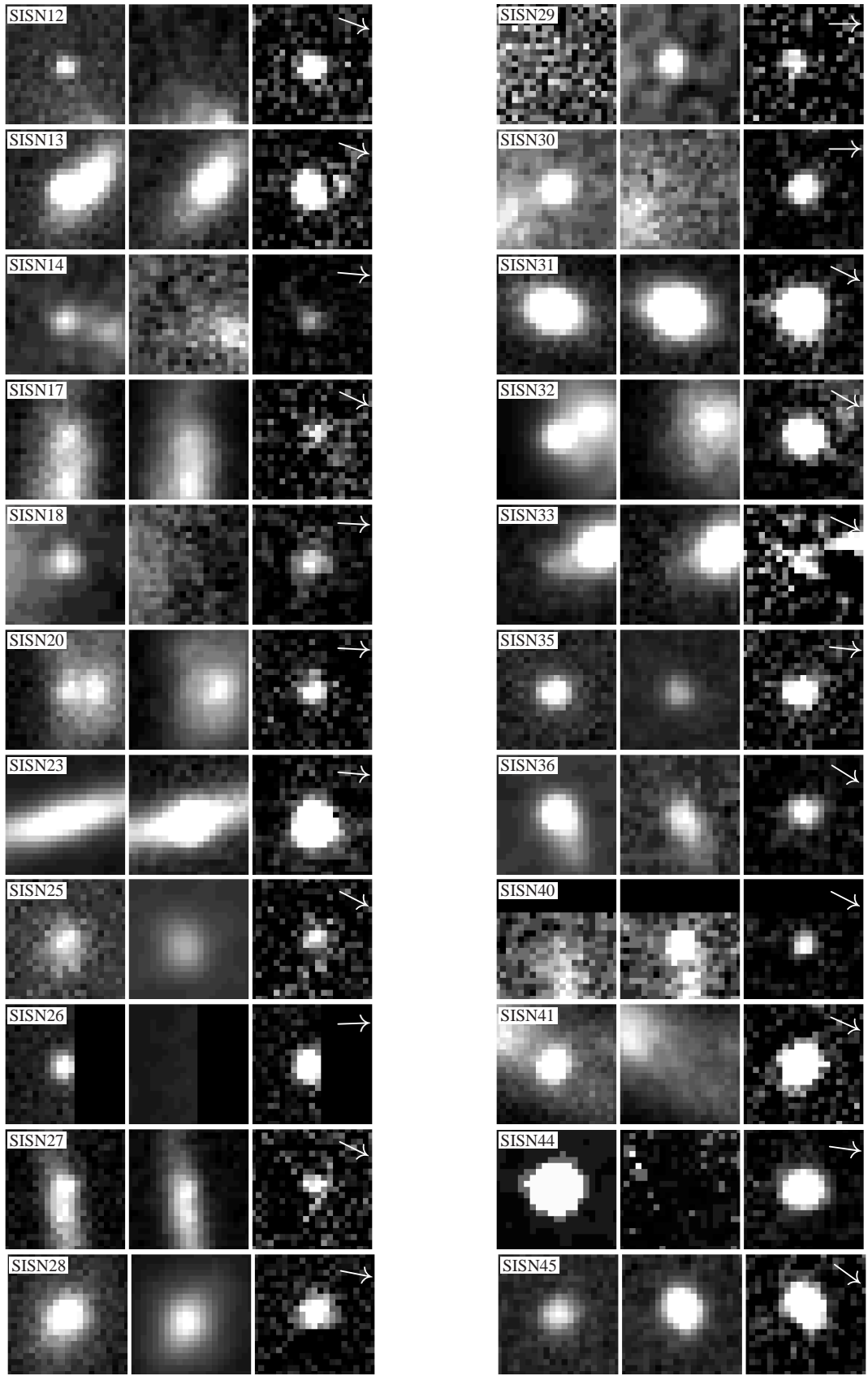

\begin{minipage}[t]{.45\textwidth}
\centering

\begin{lpic}{f4h(0.35)}
\setlength{\lpbgboxsep}{0.2mm}
\lbl[W]{15.5,51;SISN12}
\lbl{164,47,-110;{\huge{\color{white}$ \uparrow$}}}
\end{lpic}
\begin{lpic}{f4i(0.35)}
\setlength{\lpbgboxsep}{0.2mm}
\lbl[W]{15.5,51;SISN13}
\lbl{164,47,-110;{\huge{\color{white}$ \uparrow$}}}
\end{lpic}
\begin{lpic}{f4j(0.35)}
\setlength{\lpbgboxsep}{0.2mm}
\lbl[W]{15.5,51;SISN14}
\lbl{164,47,-95;{\huge{\color{white}$ \uparrow$}}}
\end{lpic}
\begin{lpic}{f4k(0.35)}
\setlength{\lpbgboxsep}{0.2mm}
\lbl[W]{15.5,51;SISN17}
\lbl{164,47,-115;{\huge{\color{white}$ \uparrow$}}}
\end{lpic}
\begin{lpic}{f4l(0.35)}
\setlength{\lpbgboxsep}{0.2mm}
\lbl[W]{15.5,51;SISN18}
\lbl{164,47,-93;{\huge{\color{white}$ \uparrow$}}}
\end{lpic}
\begin{lpic}{f4m(0.35)}
\setlength{\lpbgboxsep}{0.2mm}
\lbl[W]{15.5,51;SISN20}
\lbl{164,47,-93;{\huge{\color{white}$ \uparrow$}}}
\end{lpic}
\begin{lpic}{f4n(0.35)}
\setlength{\lpbgboxsep}{0.2mm}
\lbl[W]{15.5,51;SISN23}
\lbl{164,47,-95;{\huge{\color{white}$ \uparrow$}}}
\end{lpic}
\begin{lpic}{f4o(0.35)}
\setlength{\lpbgboxsep}{0.2mm}
\lbl[W]{15.5,51;SISN25}
\lbl{164,47,-117;{\huge{\color{white}$ \uparrow$}}}
\end{lpic}
\begin{lpic}{f4p(0.35)}
\setlength{\lpbgboxsep}{0.2mm}
\lbl[W]{15.5,51;SISN26}
\lbl{164,47,-88;{\huge{\color{white}$ \uparrow$}}}
\end{lpic}
\begin{lpic}{f4q(0.35)}
\setlength{\lpbgboxsep}{0.2mm}
\lbl[W]{15.5,51;SISN27}
\lbl{164,47,-116;{\huge{\color{white}$ \uparrow$}}}
\end{lpic}
\begin{lpic}{f4r(0.35)}
\setlength{\lpbgboxsep}{0.2mm}
\lbl[W]{15.5,51;SISN28}
\lbl{164,47,-101;{\huge{\color{white}$ \uparrow$}}}
\end{lpic}

\end{minipage}
\begin{minipage}[t]{.45\textwidth}
\centering

\begin{lpic}{f4s(0.35)}
\setlength{\lpbgboxsep}{0.2mm}
\lbl[W]{15.5,51;SISN29}
\lbl{164,47,-90;{\huge{\color{white}$ \uparrow$}}}
\end{lpic}
\begin{lpic}{f4t(0.35)}
\setlength{\lpbgboxsep}{0.2mm}
\lbl[W]{15.5,51;SISN30}
\lbl{164,47,-90;{\huge{\color{white}$ \uparrow$}}}
\end{lpic}
\begin{lpic}{f4u(0.35)}
\setlength{\lpbgboxsep}{0.2mm}
\lbl[W]{15.5,51;SISN31}
\lbl{164,47,-116;{\huge{\color{white}$ \uparrow$}}}
\end{lpic}
\begin{lpic}{f4v(0.35)}
\setlength{\lpbgboxsep}{0.2mm}
\lbl[W]{15.5,51;SISN32}
\lbl{164,47,-120;{\huge{\color{white}$ \uparrow$}}}
\end{lpic}
\begin{lpic}{f4ca(0.35)}
\setlength{\lpbgboxsep}{0.2mm}
\lbl[W]{15.5,51;SISN33}
\lbl{164,47,-115;{\huge{\color{white}$ \uparrow$}}}
\end{lpic}
\begin{lpic}{f4cb(0.35)}
\setlength{\lpbgboxsep}{0.2mm}
\lbl[W]{15.5,51;SISN35}
\lbl{164,47,-96;{\huge{\color{white}$ \uparrow$}}}
\end{lpic}
\begin{lpic}{f4cc(0.35)}
\setlength{\lpbgboxsep}{0.2mm}
\lbl[W]{15.5,51;SISN36}
\lbl{164,47,-122;{\huge{\color{white}$ \uparrow$}}}
\end{lpic}
\begin{lpic}{f4cd(0.35)}
\setlength{\lpbgboxsep}{0.2mm}
\lbl[W]{15.5,51;SISN40}
\lbl{164,47,-117;{\huge{\color{white}$ \uparrow$}}}
\end{lpic}
\begin{lpic}{f4ce(0.35)}
\setlength{\lpbgboxsep}{0.2mm}
\lbl[W]{15.5,51;SISN41}
\lbl{164,47,-114;{\huge{\color{white}$ \uparrow$}}}
\end{lpic}
\begin{lpic}{f4cf(0.35)}
\setlength{\lpbgboxsep}{0.2mm}
\lbl[W]{15.5,51;SISN44}
\lbl{164,47,-98;{\huge{\color{white}$ \uparrow$}}}
\end{lpic}
\begin{lpic}{f4cg(0.35)}
\setlength{\lpbgboxsep}{0.2mm}
\lbl[W]{15.5,51;SISN45}
\lbl{164,47,-126;{\huge{\color{white}$ \uparrow$}}}
\end{lpic}

\end{minipage}
\caption{[continued]}
\end{figure*}


\setcounter{figure}{4}
\begin{figure}
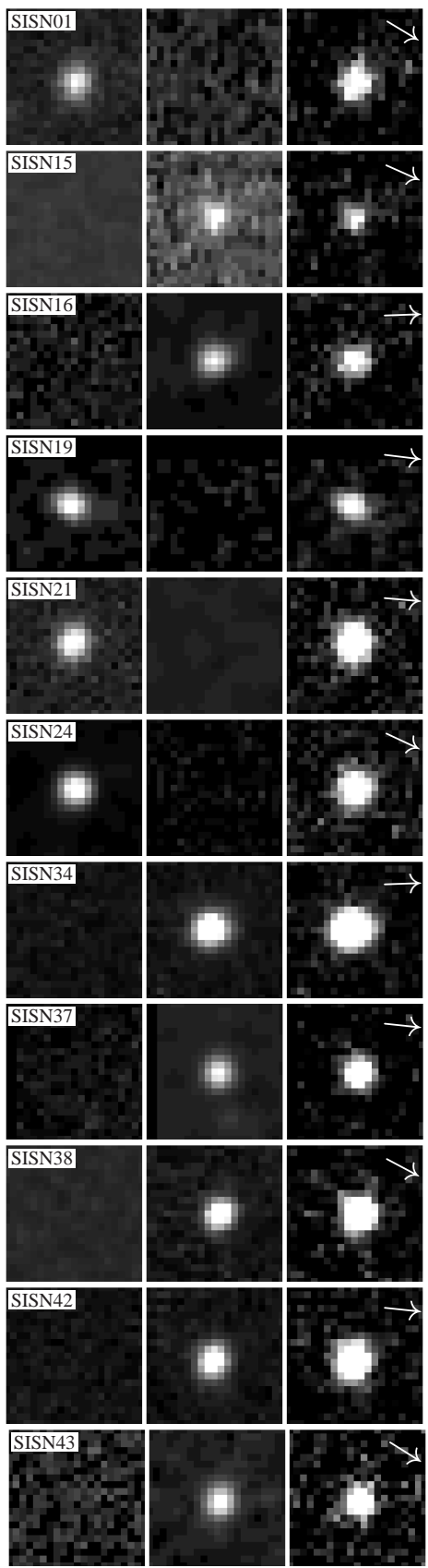

\centering
\begin{lpic}{f5a(0.35)}
\setlength{\lpbgboxsep}{0.2mm}
\lbl[W]{15.5,51;SISN01}
\lbl{164,47,-121;{\huge{\color{white}$ \uparrow$}}}
\end{lpic}
\begin{lpic}{f5b(0.35)}
\setlength{\lpbgboxsep}{0.2mm}
\lbl[W]{15.5,51;SISN15}
\lbl{164,47,-114;{\huge{\color{white}$ \uparrow$}}}
\end{lpic}
\begin{lpic}{f5c(0.35)}
\setlength{\lpbgboxsep}{0.2mm}
\lbl[W]{15.5,51;SISN16}
\lbl{164,47,-88;{\huge{\color{white}$ \uparrow$}}}
\end{lpic}
\begin{lpic}{f5d(0.35)}
\setlength{\lpbgboxsep}{0.2mm}
\lbl[W]{15.5,51;SISN19}
\lbl{164,47,-97;{\huge{\color{white}$ \uparrow$}}}
\end{lpic}
\begin{lpic}{f5e(0.35)}
\setlength{\lpbgboxsep}{0.2mm}
\lbl[W]{15.5,51;SISN21}
\lbl{164,47,-95;{\huge{\color{white}$ \uparrow$}}}
\end{lpic}
\begin{lpic}{f5f(0.35)}
\setlength{\lpbgboxsep}{0.2mm}
\lbl[W]{15.5,51;SISN24}
\lbl{164,47,-115;{\huge{\color{white}$ \uparrow$}}}
\end{lpic}
\begin{lpic}{f5g(0.35)}
\setlength{\lpbgboxsep}{0.2mm}
\lbl[W]{15.5,51;SISN34}
\lbl{164,47,-88;{\huge{\color{white}$ \uparrow$}}}
\end{lpic}
\begin{lpic}{f5h(0.35)}
\setlength{\lpbgboxsep}{0.2mm}
\lbl[W]{15.5,51;SISN37}
\lbl{164,47,-96;{\huge{\color{white}$ \uparrow$}}}
\end{lpic}
\begin{lpic}{f5i(0.35)}
\setlength{\lpbgboxsep}{0.2mm}
\lbl[W]{15.5,51;SISN38}
\lbl{164,47,-118;{\huge{\color{white}$ \uparrow$}}}
\end{lpic}
\begin{lpic}{f5j(0.35)}
\setlength{\lpbgboxsep}{0.2mm}
\lbl[W]{15.5,51;SISN42}
\lbl{164,47,-96;{\huge{\color{white}$ \uparrow$}}}
\end{lpic}
\begin{lpic}{f5k(0.35)}
\setlength{\lpbgboxsep}{0.2mm}
\lbl[W]{15.5,51;SISN43}
\lbl{164,47,-122;{\huge{\color{white}$ \uparrow$}}}
\end{lpic}

\caption{Same as Fig. 4 but for the hostless candidate sample.}
\end{figure}

\setcounter{figure}{5}
\begin{figure}
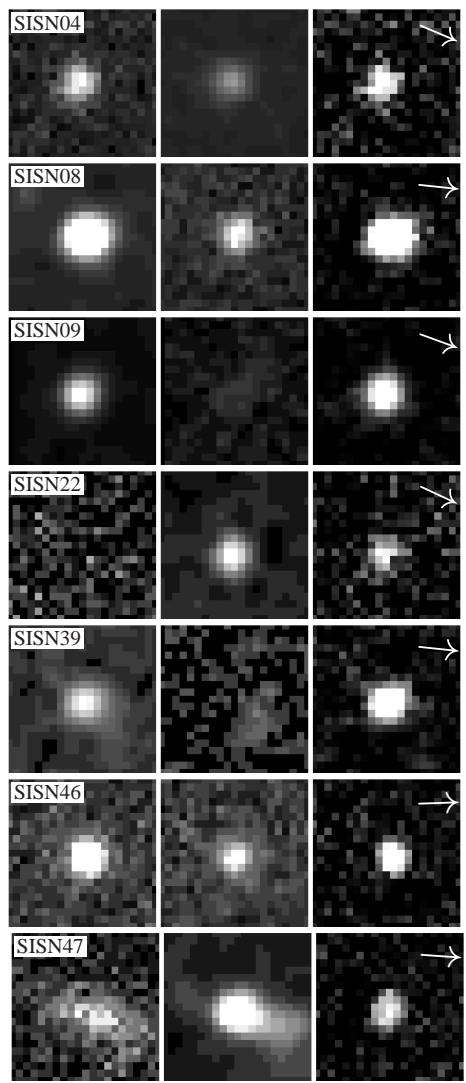

\centering

\begin{lpic}{f6a(0.35)}
\setlength{\lpbgboxsep}{0.2mm}
\lbl[W]{15.5,51;SISN04}
\lbl{164,47,-113;{\huge{\color{white}$ \uparrow$}}}
\end{lpic}
\begin{lpic}{f6b(0.35)}
\setlength{\lpbgboxsep}{0.2mm}
\lbl[W]{15.5,51;SISN08}
\lbl{164,47,-96;{\huge{\color{white}$ \uparrow$}}}
\end{lpic}
\begin{lpic}{f6c(0.35)}
\setlength{\lpbgboxsep}{0.2mm}
\lbl[W]{15.5,51;SISN09}
\lbl{164,47,-110;{\huge{\color{white}$ \uparrow$}}}
\end{lpic}
\begin{lpic}{f6d(0.35)}
\setlength{\lpbgboxsep}{0.2mm}
\lbl[W]{15.5,51;SISN22}
\lbl{164,47,-115;{\huge{\color{white}$ \uparrow$}}}
\end{lpic}
\begin{lpic}{f6e(0.35)}
\setlength{\lpbgboxsep}{0.2mm}
\lbl[W]{15.5,51;SISN39}
\lbl{164,47,-96;{\huge{\color{white}$ \uparrow$}}}
\end{lpic}
\begin{lpic}{f6f(0.35)}
\setlength{\lpbgboxsep}{0.2mm}
\lbl[W]{15.5,51;SISN46}
\lbl{164,47,-88;{\huge{\color{white}$ \uparrow$}}}
\end{lpic}
\begin{lpic}{f6g(0.35)}
\setlength{\lpbgboxsep}{0.2mm}
\lbl[W]{15.5,51;SISN47}
\lbl{164,47,-94;{\huge{\color{white}$ \uparrow$}}}
\end{lpic}

\caption{Same as Fig. 4 but for the high-$z$ candidate sample, i.e. the
  seven candidates rejected based on
  their high photometric host redshifts.}
\end{figure}


\subsection{Supernova classification}

Spectroscopic classification of our SN candidates is, of course,
impossible in this archival survey. We therefore adopt a photometric
classification method, the Supernova Automated Bayesian Classification
(SN-ABC) routine of Poznanski, Maoz, \& Gal-Yam (2007a). This method
compares the SN candidate magnitudes to a sample of SN spectral
templates of different types, ages, redshifts, and extinctions and
returns the probability of a candidate being a Type Ia SN, $P({\rm
  Ia})$, as opposed to being a core-collapse SN. The routine uses as a
prior the host redshift probability distribution function of each
candidate. Six candidates in our sample have spectroscopic host
redshifts. For the rest of the candidates we retrieve photometric
redshifts from the SDSS catalog. We then assume Gaussian redshift
probability distribution functions using the host spectroscopic or
photometric redshifts and their errors.


We next perform an analysis similar to that of Poznanski et al.
(2007b) in order to determine the degree of classification uncertainty
and bias, i.e., what is the most probable real range in the fraction
of SNe Ia, given the fraction that is classified as such. Using the SN
spectral templates, we create a sample of fake SNe of the four most
prevalent types: Ia, IIP, IIn, and Ib/c. The distribution of redshifts
is according to the redshift distribution of the galaxy sample used in
the detection efficiency simulations (see $\S4$).  We note that while
Type Ia's and IIn's are observable up to a redshift of $z\approx
0.35$, given the SDSS limiting magnitude, types Ib/c and IIP are
observable only out to $z\approx 0.15$. We find that, in our redshift
range, the value of $P({\rm Ia})$ is not indicative of the actual
probability of an object being a SN Ia. This is due to the fact that
while SNe Ia usually are assigned very high values ($85\%$ get $P({\rm
  Ia})>0.9$) for core-collapse SNe the results are less obvious. About
half the type IIn and Ib/c SNe, and a fifth of the type IIP SNe get
misclassified, with values of $P({\rm Ia})>0.5$. As a consequence we use
$P({\rm Ia})$ as a quality indicator, rather than a probability,
cutting the sample at $P({\rm Ia})>0.9$ where the contamination by
core-collapse SNe is minimal. The fractions of simulated SNe that are
classified or misclassified as Type Ia by the SN-ABC routine using the
above threshold are $0.85,~0.04,~0.28$, and $0.13$ for types Ia, IIP,
IIn, and Ib/c, respectively. Applying SN-ABC together with a $P({\rm
  Ia})>0.9$ threshold on the $29$ candidates in our sample results in
$16$ candidates which are classified as Ia's, with $\chi^{2}$ values
in the range $0.3-10.6$. Table $2$ lists the candidate properties,
including SN types and host redshifts.

Since we do not know the real fraction of Type Ia SNe at low redshift,
we create a sample of possible fractions of the different types, e.g.,
$30\%~{\rm Ia}, ~50\%~{\rm IIP},~ 10\%~{\rm IIn},~ 10\%~{\rm Ib/c}$,
using steps of $10\%$. For each set of fractions, we calculate the
binomial probability of finding the observed $16$ SNe Ia out of $29$,
using the fractions of SNe classified as Type Ia that we calculated
for the four SN types. Each of the combinations is given a weight
according to the number of permutations with the same fraction of SNe
Ia. The outcome of this calculation is a probability distribution for
the real number of SNe in our sample, given that $16$ SNe were
classified as Type Ia.  From this probability function we derive that
the most probable ``true'' number of SNe Ia in our sample is $N_{{\rm
    Ia}}=17^{+3.8}_{-3.1}\pm 3.8$, where the first error is the $68\%$
range Poisson error, and the second one is a systematic $68\%$ range
error due to the uncertainty in classification.
\begin{table*}
\scriptsize
\begin{center}
\caption{SN candidate sample}
\smallskip
\begin{tabular}{ccccccccccccc}
\hline
\noalign{\smallskip}
Id & RA & Dec & $\beta ({\rm deg})$ & $u$ & $g$ & $r$ & $i$ & $z$ &
$E(B-V)$ & Redshift & $P(\rm{Ia})$ & Sample\\
\hline
\noalign{\smallskip}
SISN01 & 15:44:10.57 & +51:46:30.57 & 67.7 & 22.43 & 20.72 & 20.91 & 21.04 &
---- & 0.014 & ---- & 0.98 & hostless \\
SISN02 & 15:45:04.35 & +40:11:38.27 & 57.7 & 20.94 & 20.95 & 21.54 & 21.80 &
21.65 & 0.016 & 0.13 & 0.02 & main \\
\bf SISN03 & 14:56:43.00 & +42:29:17.73 & 55.5 & 22.40 & 21.47 & 21.49 & 21.22 &
21.59 & 0.019 & 0.20 & 0.98 & main\\
SISN04 & 14:57:06.28 & +46:37:25.51 & 59.0 & 21.61 & 21.31 & 21.64 & 21.07 &
22.38 & 0.015 & 0.61 & 0.42 & high-z\\
\bf SISN05 & 14:56:48.96 & +10:57:47.53 & 26.5 & 22.84 & 21.18 & 21.24 & 21.18 &
21.41 & 0.034 & 0.16 & 0.99 & main\\
\bf SISN06 & 14:56:49.05 & +36:36:56.41 & 50.3 & 22.05 & 20.36 & 20.56 & 20.74 &
21.65 & 0.015 & 0.20 & 1.00 & main\\
\bf SISN07 & 14:42:59.88 & +62:42:48.38 & 69.1 & 22.44 & 20.59 & 20.39 & 20.69 &
20.65 & 0.015 & 0.16 & 0.98 & main\\
SISN08 & 14:44:51.65 & +06:55:48.55 & 21.8 & 20.36 & 20.39 & 20.50 & 20.71 &
21.25 & 0.033 & 0.76 & 0.01 & high-z\\
SISN09 & 14:46:19.33 & +53:47:14.39 & 63.7 & 22.66 & 21.04 & 20.87 & 21.60 &
21.22 & 0.010 & 0.65 & 0.93 & high-z\\
SISN10 & 14:49:18.82 & +52:33:07.73 & 63.0 & 23.66 & 21.54 & 21.35 & 21.66 &
21.23 & 0.015 & {\bf 0.07} & 0.02 & main\\
\bf SISN11 & 15:00:35.10 & +00:48:53.50 & 17.2 & 23.63 & 20.91 & 20.97 & 20.74 &
21.69 & 0.049 & 0.20 & 0.99 & main\\
SISN12 & 15:00:58.34 & +29:04:11.96 & 43.8 & 22.57 & 20.98 & 20.50 & 20.89 &
21.25 & 0.018 & 0.04 & 0.19 & main\\
\bf SISN13 & 15:02:35.07 & +28:56:59.42 & 43.9 & 21.79 & 19.72 & 19.30 & 19.79 &
20.07 & 0.023 & 0.14 & 1.00 & main\\
SISN14 & 15:03:47.16 & +04:56:04.76 & 21.3 & ---- & 21.96 & 20.97 & 21.19 &
20.82 & 0.040 & 0.44 & 0.39 & main\\
SISN15 & 15:07:33.29 & +45:33:49.61 & 59.1 & 22.21 & 21.75 & 21.31 & 21.29 &
21.41 & 0.021 & ---- & 0.23 & hostless\\
SISN16 & 15:08:45.09 & -00:38:53.27 & 16.3 & 23.41 & 21.07 & 20.51 & 20.33 &
21.06 & 0.065 & ---- & 0.74 & hostless\\
\bf SISN17 & 15:08:37.26 & +45:29:16.72 & 59.2 & 22.98 & 22.20 & 21.31 & 21.06 &
21.31 & 0.025 & 0.09 & 0.93 & main\\
\bf SISN18 & 15:12:23.54 & +02:40:28.03 & 19.8 & 22.44 & 21.18 & 21.15 & 21.01 &
---- & 0.041 & 0.23 & 1.00 & main\\
SISN19 & 15:12:44.37 & +06:20:37.23 & 23.3 & 21.95 & 21.52 & 20.97 & 20.65 &
20.17 & 0.034 & ---- & 0.57 & hostless\\
SISN20 & 15:14:43.66 & +04:39:55.11 & 21.9 & ---- & 21.31 & 21.12 & 20.91 &
20.81 & 0.051 & 0.10 & 0.19 & main\\
SISN21 & 15:16:38.95 & +05:57:26.38 & 23.2 & 20.82 & 20.13 & 19.90 & 19.78 &
19.87 & 0.042 & ---- & 0.35 & hostless\\
SISN22 & 15:16:56.21 & +47:10:04.82 & 61.4 & 21.10 & 21.47 & 20.86 & 21.22 &
22.26 & 0.030 & 0.94 & 0.00 & high-z\\
SISN23 & 15:17:32.35 & +04:30:02.51 & 21.9 & 21.22 & 19.43 & 19.31 & 19.55 &
18.72 & 0.047 & 0.00 & 0.08 & main\\
SISN24 & 15:17:33.01 & +39:19:52.54 & 54.6 & 22.24 & 20.81 & 20.71 & 21.09 &
21.36 & 0.017 & ---- & 0.95 & hostless\\
\bf SISN25 & 15:22:29.67 & +38:46:35.4 & 54.5 & 22.67 & 21.68 & 21.56 & 21.36 &
21.72 & 0.017 & 0.25 & 0.99 & main\\
\bf SISN26 & 15:33:57.56 & -00:48:42.1 & 17.9 & 21.50 & 20.27 & 20.25 & 20.48 &
21.16 & 0.116 & {\bf 0.12} & 1.00 & main\\
\bf SISN27 & 15:33:07.47 & +29:54:54.7 & 47.2 & 23.18 & 22.15 & 21.03 & 21.22 &
21.19 & 0.03 & 0.08 & 0.95 & main\\
SISN28 & 15:33:40.44 & +09:16:37.1 & 27.6 & 20.62 & 20.57 & 20.72 & 20.59 &
20.96 & 0.041 & 0.05 & 0.02 & main\\
\bf SISN29 & 15:37:37.48 & -00:38:37.3 & 18.2 & 22.25 & 22.32 & 21.02 & 20.90 &
20.97 & 0.098 & {\bf 0.16} & 1.00 & main\\
SISN30 & 15:43:11.48 & -00:23:55.6 & 18.8 & 21.70 & 20.70 & 20.38 & 20.14 &
19.96 & 0.096 & 0.04 & 0.04 & main\\
SISN31 & 15:45:27.58 & +26:27:58.8 & 44.9 & 19.88 & 18.61 & 18.07 & 18.16 &
17.82 & 0.049 & {\bf 0.03} & 0.01 & main\\
\bf SISN32 & 15:45:46.02 & +35:37:06.7 & 53.4 & 21.73 & 20.02 & 19.55 & 19.98 &
20.59 & 0.029 & 0.06 & 0.97 & main\\
\bf SISN33 & 15:46:03.12 & +22:58:53.0 & 41.6 & 23.01 & 22.47 & 21.17 & 21.21 &
20.94 & 0.055 & {\bf 0.12} & 1.00 & main\\
SISN34 & 15:46:15.29 & -00:37:05.8 & 18.8 & 21.76 & 19.83 & 19.25 & 18.91 &
18.94 & 0.103 & ---- & 0.06 & hostless\\
SISN35 & 15:46:48.36 & +03:26:30.7 & 22.8 & 20.68 & 20.30 & 20.83 & 20.76 &
20.89 & 0.094 & 0.10 & 0.83 & main\\
SISN36 & 15:49:35.40 & +39:59:10.9 & 57.8 & 22.28 & 21.17 & 20.49 & 20.15 &
20.55 & 0.012 & 0.56 & 0.86 & main\\
SISN37 & 15:51:22.86 & +04:19:46.6 & 23.9 & 21.33 & 20.71 & 20.65 & 20.89 &
22.35 & 0.079 & ---- & 0.96 & hostless\\
SISN38 & 15:52:32.30 & +25:38:47.8 & 44.6 & 22.58 & 20.58 & 20.02 & 20.35 &
21.72 & 0.060 & ---- & 0.99 & hostless\\
SISN39 & 15:52:55.62 & +03:40:15.3 & 23.3 & 23.30 & 20.55 & 20.27 & 20.62 &
---- & 0.153 & 1.18 & 0.97 & high-z\\
SISN40 & 15:54:12.31 & +24:15:30.8 & 43.3 & 23.52 & 21.85 & 21.11 & 21.24 &
22.42 & 0.048 & 0.08 & 0.59 & main\\
SISN41 & 15:54:52.01 & +21:07:10.8 & 40.4 & 24.77 & 19.03 & 17.90 & 17.75 &
17.48 & 0.055 & 0.04 & 0.00 & main\\
SISN42 & 15:54:26.70 & +03:41:09.8 & 23.4 & 21.85 & 20.00 & 19.65 & 19.57 &
19.69 & 0.154 & ---- & 0.64 & hostless\\
SISN43 & 15:55:53.55 & +31:23:24.4 & 50.3 & 21.44 & 21.08 & 21.21 & 21.16 &
21.10 & 0.025 & ---- & 0.34 & hostless\\
\bf SISN44 & 15:58:38.83 & +05:15:48.3 & 25.2 & 22.12 & 19.95 & 19.22 & 19.84 &
20.05 & 0.056 & {\bf 0.07} & 1.00 & main\\
\bf SISN45 & 15:59:11.36 & +46:17:49.3 & 64.3 & 21.66 & 19.75 & 19.39 & 19.97 &
20.94 & 0.016 & 0.15 & 1.00 & main\\
SISN46 & 15:43:30.40 & -01:11:51.3 & 18.1 & 21.64 & 21.00 & 21.52 & 21.85 &
21.05 & 0.117 & 0.56 & 0.79 & high-z\\
SISN47 & 15:39:04.47 & +03:48:51.4 & 22.7 & 21.73 & 21.22 & 20.98 & 21.04 &
20.20 & 0.061 & 0.51 & 0.04 & high-z\\
\noalign{\smallskip}
\hline
\smallskip
\end{tabular}
\end{center}
Notes: Candidates with names in boldface are classified as
Type Ia in the final sample.
Magnitudes are before correction for Galactic
extinction, derived from the reddening listed in the $E(B-V)$
column. 
Photometric errors are
according to Table 1. Redshifts in boldface are
spectroscopic. Candidates with no redshift are apparently
hostless. The $P(Ia)$ values of the candidates in the hostless and high-$z$
samples were calculated assuming a uniform redshift probability
distribution in the range $0<z<0.35$. $\beta$ is the ecliptic latitude.
\end{table*}

\subsection{Supernova rate}

We now derive the luminosity-normalized SN Ia rate, $r_{{\rm Ia}}$, in SNu$_{{\rm band}}$
units\footnote{1 SNu$_{band}=$ SN $(100~{\rm
    yr}~10^{10}{\rm L}^{band}_{\sun})^{-1}$}. The rate
is calculated using
\begin{equation}
r_{{\rm Ia}}=\frac{N_{{\rm Ia}}}{\displaystyle \sum_{i} \int{\eta_{i}(z)j_{{\rm lim}}(z)dV}},
\end{equation}
where 
$N_{{\rm Ia}}$ is the number of SNe Ia, 
$dV$ is a comoving volume element, $j_{{\rm lim}}$ is the
luminosity density originating from galaxies which are above the SDSS
limiting magnitude, and $\eta_{i}(z)$ is the effective visibility time
(or ``control time'') of the
$i$-th image set, i.e., the time during which the SN is detectable.
The integration is over the cosmological volume in each set, and the
summation is over image sets. Given $\epsilon (m)$, the detection
efficiency function as a function of magnitude $m$,
\begin{equation}
\eta_{i} (z)= \int{\epsilon[m_{{\rm eff}}(t)]\frac{dm_{{\rm eff}}}{dt}dt},
\end{equation}
where, $m_{{\rm eff}}(t)$ is the effective SN light curve determined by the
time difference, $\Delta t_{i}$, between the reference and registered images in each
set $i$,
\begin{equation}
m_{{\rm eff}}(t)=-2.5{\rm log}\left(10^{-0.4m(t)}-10^{-0.4m(t+\Delta
t_{i})}\right).
\end{equation}

The mean redshift which we probe in this work, which depends on the
visibility time and thus on the efficiency function, is given by
\begin{equation}
\left<z\right>=\frac{\int{\eta(z)z~\frac{dV}{dz}dz}}{\int{\eta(z)\frac{dV}{dz}dz}}.
\end{equation}

In order to calculate the luminosity density, $j(z)$, we again use the
Blanton et al. (2003) galaxy luminosity function. We convert their
luminosity function, which is given for the SDSS bands shifted to
$z=0.1$, back to the rest-frame SDSS bands. We also account for
luminosity evolution using their luminosity evolution parameter $Q$,
thus obtaining the luminosity density as a function of redshift.
Integrating over the luminosity function up to the limiting magnitude
at each redshift provides $j_{{\rm lim}}(z)$.

The SN Ia rate must also be corrected for host extinction. Riello \&
Patat (2005) performed Monte Carlo simulations in which they modeled
the dust distribution in host galaxies and accounted for various
bulge-to-disc ratios and total optical depths. They found that the
factor, $f$, by which SN Ia rates need to be corrected, is
$1.27<f<1.91$, for Milky-Way-like dust. A similar, though lower,
correction factor was derived by Neill et al. (2006), who derived the
type Ia SN rate at $z\approx0.5$. Neill et al. (2006) considered both
Gaussian and exponential host extinction distributions in their
detection efficiency simulations. They found a correction factor of
$1.10<f<1.37$. Based on these studies, we adopt an intermediate
correction factor of $f=1.25$ to our SN rates.

The derived SN Ia rates in the
$g$ and $r$ bands, using Eq. $2$ and also correcting for host
extinction, are
\begin{eqnarray}
r^{r}_{{\rm Ia}}&=&\left(11.5^{+2.5+1.1}_{-2.5-0.9}\pm 2.5\right)\times
10^{-2}~h_{70}^{2}~{\rm SNu}_{r}, \nonumber \\
r^{g}_{{\rm Ia}}&=&\left(14.0^{+2.5+1.4}_{-2.5-1.1}\pm 2.5\right)\times
10^{-2}~h_{70}^{2}~{\rm SNu}_{g} \nonumber
\end{eqnarray}
at a mean redshift of $\left<z\right>=0.20$. The first error is due to
the Poisson fluctuations in the SN number. The second is a systematic
error due to the uncertainty in the detection
efficiency function, calculated by
using the efficiency function upper and lower $(1\sigma)$ limits. The
third error is the systematic classification error.

For comparison with previously published rates, we convert our
luminosity-normalized rates also to a volumetric rate. We do so by
replacing Eq. 2 with
\begin{equation}
R_{{\rm Ia}}=\frac{N_{{\rm Ia}}}{\displaystyle \sum_{i} \int{\eta_{i}(z)\frac{j_{{\rm
          lim}}(z)}{j_{{\rm total}}(z)}dV}},
\end{equation}
where $j_{{\rm total}}$ is the total luminosity density. The resultant
volumetric rates are then
\begin{eqnarray}
R^{r}_{Ia}(0.2)&=&\left(1.75^{+0.40+0.17}_{-0.32-0.14}\pm 0.40\right)\times
10^{-5}~{\rm yr}^{-1}~h_{70}^{3}~{\rm Mpc}^{-3}, \nonumber \\
R^{g}_{Ia}(0.2)&=&\left(1.89^{+0.42+0.18}_{-0.34-0.15}\pm
  0.42\right)\times 10^{-5}~{\rm yr}^{-1}~h_{70}^{3}~{\rm Mpc}^{-3}. \nonumber
\end{eqnarray}
The difference between these two volumetric rates is due to the
different evolution of the luminosity density in each band. This
difference is an inherent weakness of deriving volumetric rates from
luminosity-normalized rates, but in our case the difference is smaller
than any of the other sources of uncertainty.

\section{Comparison with previous measurements}

In this section, we compare our rate measurements to previously
reported low-redshift SN rates. Most of these measurements (e.g.,
Cappellaro et al. 1999; Hardin et al. 2000; Blanc et al. 2004) were
given in $B$-band SNu units. These rates were then
converted to volumetric rates using the luminosity density at the
relevant redshift. However, various luminosity functions were used to
convert to volumetric rates. For example, Blanc et al. converted their
rate and the rates of Cappellaro et al.  (1999), Hardin et al.
(2000), and Madgwick et al. (2003), using the 2dF redshift survey
luminosity density (Cross et al. 2001). In contrast, Botticella et al.
(2008) fit a set of luminosity density measurements (Norberg et al.
2002; Bell et al.  2004; Blanton et al.  2003; Faber et al. 2007;
Tresse et al.  2007), with a smooth function of redshift, and used it
to perform the conversion to volumetric rates. We will repeat the Botticella et al. (2008) conversion of volumetric rates of previously published
luminosity-normalized rates (see Table 3), using their redshift-dependent luminosity
density,
\begin{equation}
j_{B}(z)=(1.03+1.76\times z)\times 10^{8}~L_{\sun}^{B}~{\rm Mpc}^{-3}.
\end{equation}

Dilday et al. (2008) have recently reported a Type Ia SN rate
from the SDSS-II Supernova Survey (Frieman et al. 2008). The majority
of SNe found in this survey, in contrast to our survey, have been
confirmed spectroscopically (resulting in a lower redshift range being
probed). The SN Ia rate measured by Dilday et al.,
based on $17$ SNe at $z\sim 0.09$ is higher than our rate measurement
by a factor of $\sim 1.5$ but consistent within the errors. Table 3
shows these various low-redshift rate measurements and Fig. $7$ shows a compilation of rate measurements to $z<0.5$.
\begin{table*}

\begin{minipage}{\textwidth}
\begin{center}
\caption{Comparison of low-redshift Type Ia SN rate measurements.}
\begin{tabular}{@{}ccllll@{}}
\hline
\noalign{\smallskip}
$<z>$ & $N_{{\rm Ia}}$ & \multicolumn{3}{c}{$R_{Ia}$} & Author \\
\cline{3-5}
\noalign{\smallskip}
     & & $h_{70}^{2}~{\rm SNu}_{B}$ & $h_{70}^{2}~{\rm SNu}_{g}$ & $10^{-5}~{\rm yr}^{-1}~h_{70}^{3}~{\rm
       Mpc}^{-3}$ &  \\
\hline
 $\sim0$ & $70$ & $0.18\pm 0.05$ & & $1.85\pm 0.5$ & Cappellaro et
 al. (1999)$^b$ \\
 \noalign{\smallskip}
 $0.09$ & $17$ & & $0.235^{+0.07}_{-0.06}$ & $2.9^{+0.9}_{-0.7}$ & Dilday et al. (2008)$^a$ \\
 \noalign{\smallskip}
 $0.098$ & $19$ & $0.196\pm 0.098$ & & $2.4\pm 1.2$ & Madgwick et al. (2003)$^b$ \\
 \noalign{\smallskip}
 $0.13$ & $14$ & $0.125^{+0.044+0.028}_{-0.034-0.028}$ & &
 $1.58^{+0.56+0.35}_{-0.43-0.35}$ & Blanc et al. (2004)$^b$ \\
 \noalign{\smallskip}
 $0.14$ & $4$ & $0.22^{+0.17+0.06}_{-0.10-0.03}$ & &
 $2.8^{+2.2+0.7}_{-1.3-0.4}$ & Hardin et al. (2000)$^b$ \\
 \noalign{\smallskip}
 ${\bf 0.2}$ & ${\bf 17}$ & ${\bf 0.14^{+0.03+0.01}_{-0.03-0.01}\pm 0.03}$ & ${\bf 0.14^{+0.03+0.01}_{-0.03-0.01}\pm 0.03}$ &
 ${\bf 1.89^{+0.42+0.18}_{-0.34-0.15}\pm 0.42}$ & {\bf This work}  \\
 \noalign{\smallskip}
 $0.25$ & $1$ & & & $1.7\pm 1.7$ & Barris \& Tonry (2006) \\

\noalign{\smallskip}
\hline
\end{tabular}
\end{center}
\end{minipage}

  Notes: $^a$ Luminosity-normalized rate derived from a
  volumetric rate. $^b$ These rates have been converted to volumetric
  rates using the redshift-dependent luminosity density function from Botticella et al. (2008).

\end{table*}

\setcounter{figure}{6}
\begin{figure}
\includegraphics[width=0.35\textwidth, angle=-90]{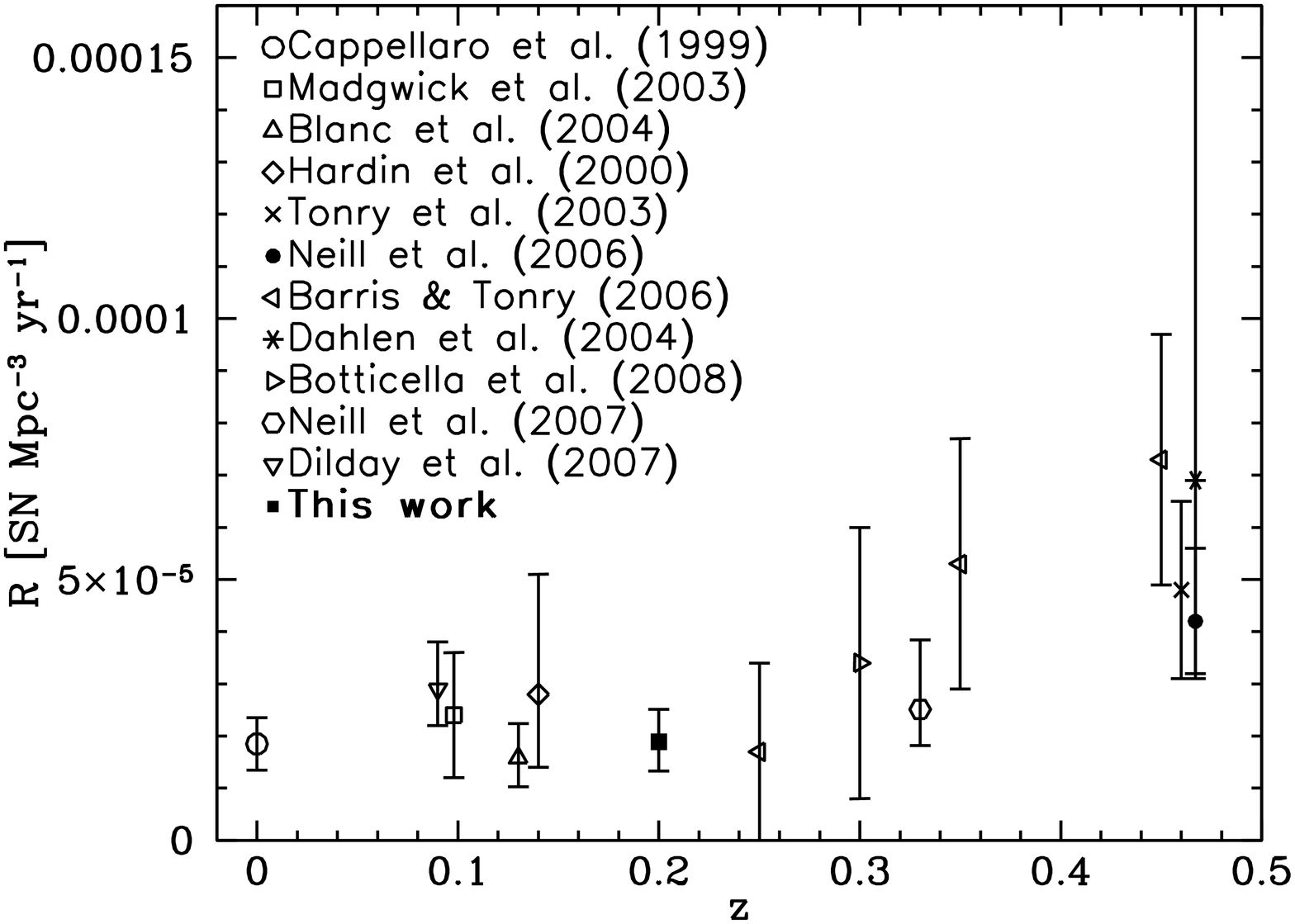}
\caption{Type Ia SN rates from different authors
  (see legend). Error bars are from quadrature additions of the various
errors listed in Table~3.}
\end{figure}

Our rate measurement is consistent with other low-redshift rate
measurements, and ranks with the most accurate among them. 
However, as mentioned
above, some of the previously published volumetric rates, which appear
in Table 3, would change, depending on the luminosity density that is
used to derive them. 
A remaining uncertainty in such comparison arises from the fact that
the rates in SNu of Cappellaro et al. (1999),
Madgwick et al. (2003), Blanc et al. (2003), and Hardin et al. (2000) are multiplied by the luminosity density at the {\it
  mean} redshift of each survey, thus not taking into account that the
effective volume of the survey is a function of redshift.

In order to compare our luminosity-normalized rate in SNu$_{g}$ units
also to rates given in SNu$_{B}$ units, we adopt the Lupton et al. (2005)
conversion between $B$ and $g$ band magnitudes,
\begin{equation}
B=g+0.2271+0.313(g-r).
\end{equation}
Using the mean luminosity-weighted color, $g-r=0.53$, of the SDSS
galaxy subset sample described in \S4, together with Eq. 8, implies a factor
of $1.03$ increase going from ${\rm SNu}_{g}$ units to ${\rm SNu}_{B}$
units. Using this factor, our rate measurement in ${\rm SNu}_{B}$
units is $r_{{\rm Ia}}=\left(14.4^{+2.6+1.4}_{-2.6-1.1}\pm 2.6\right)
\times 10^{-2}~h_{70}^{2}~{\rm SNu}_{B}$. As seen in Table~3, this
again agrees with previous measurements.

\section{Summary}

We have conducted a low-redshift photometric SN survey using archival
data from SDSS-I overlapping fields.  Based on the number of Type Ia
SNe that we find, $N_{{\rm Ia}}=17^{+3.8}_{-3.1}\pm 3.8$, and keeping
track of the various sources of error and bias, we have derived a SN
Ia rate of $r^{g}_{{\rm Ia}} = \left(14.0^{+2.5+1.4}_{-2.5-1.1}\pm
  2.5\right)\times 10^{-2}~h_{70}^{2}~{\rm SNu}_{g}$, or a volumetric
rate of $R^{g}_{{\rm Ia}} = \left(1.89^{+0.42+0.18}_{-0.34-0.15}\pm
  0.42\right)\times 10^{-5}~{\rm yr}^{-1}~h_{70}^{3}~{\rm Mpc}^{-3}$.
Our derived rates are consistent with previous measurements, but rank
with the most accurate ones. However, SN Ia rates at low-redshift,
including ours, still suffer from several sources of uncertainty. In
our case, the uncertainty is mostly due to small numbers. The
derivation of volumetric rates using different luminosity functions
and different extinction corrections is another source of ambiguity
when comparing different measurements at similar redshifts, and when
comparing observations with model predictions.  Nevertheless, we have
shown that there is a vast amount of archival SDSS data that can be
used for studying SNe at a low cost.  The full SDSS-I SN sample, once
mined, would include several hundreds of SNe, comparable to the $\sim
500$ expected from SDSS-II, (a survey designed specifically for
finding SNe). Assuming a similar fraction of Type Ia SNe as we found,
both the Poisson and binomial classification uncertainties for such a
large sample would be reduced to the $\sim 5\%$ level, while the
uncertainty due to the detection efficiency function will remain the
same.  Although SNe found by archival search methods, such as ours,
can not be studied spectroscopically, a full SDSS-I sample could be
useful for improved investigations of SN rates as a function of galaxy
type and environment.

\section*{Acknowledgments}
Funding for the SDSS and SDSS-II has been provided by the Alfred P.
Sloan Foundation, the Participating Institutions, the National Science
Foundation, the U.S. Department of Energy, the National Aeronautics
and Space Administration, the Japanese Monbukagakusho, the Max Planck
Society, and the Higher Education Funding Council for England. The
SDSS Web Site is http://www.sdss.org/. We thank the anonymous referee
for comments that improved the presentation.

\end{document}